\documentclass{aastex}

\usepackage{emulateapj5}

\shorttitle{Stars lensed by the supermassive black hole}
\shortauthors{Micha{\l}owski \& Mr{\'o}z}
\newcommand{\myemail}{mj.michalowski@gmail.com}
\newcommand{\urltt}[1]{\url{\texttt{#1}}}

\usepackage{color}

\newcommand{\msun}{\mbox{$M_\odot$}}

\newcommand{\inst}[1]{\altaffilmark{#1}}
\usepackage{comment}

\newcommand{\kms}{\mbox{km\,s$^{-1}$}}
\newcommand{\er}{\mbox{$\theta_{\rm E}$}}
\newcommand{\dls}{\mbox{$D_{\rm LS}$}}
\newcommand{\dl}{\mbox{$D_{\rm L}$}}
\newcommand{\ds}{\mbox{$D_{\rm S}$}}
\newcommand{\aksmbh}{A_{K,{\rm SMBH}}}

\begin{document}

\title{Stars lensed by the supermassive black hole in the center of the Milky Way: predictions for ELT, TMT, GMT, and JWST
}

\author{Micha{\l}~J.~Micha{\l}owski\inst{\ref{inst:uam}} \&
Przemek Mr{\'o}z\inst{\ref{inst:caltech}}
	}

\altaffiltext{1}
{Astronomical Observatory Institute, Faculty of Physics, Adam Mickiewicz University, ul.~S{\l}oneczna 36, 60-286 Pozna{\'n}, Poland \myemail  \label{inst:uam}}
\altaffiltext{2}
{Division of Physics, Mathematics, and Astronomy, California Institute of Technology, Pasadena, CA 91125, USA   \label{inst:caltech}}


\begin{abstract}
Gravitational lensing is an important prediction of general relativity, providing both its test theory and a tool to detect faint but amplified sources and to measure masses of lenses. 
For some applications, (e.g.~testing the theory), a point source lensed by a point-like lens would be more advantageous. However, until now only one gravitationally lensed star has been resolved.
The future telescopes will resolve very small  lensing  signatures for stars orbiting the supermassive black hole (SMBH) in the center of the Milky Way. 
The lensing signatures should however be easier to detect for background stars. 
We predict that ELT, TMT, and GMT will resolve the lensed images of around 100 (60) stars in the disk and 30 (20) stars in the bulge in the background of the SMBH, down to 28 (27)\,mag (Vega) limits at $K$-band, requiring 5 (1)\,hr of integration. 
In order to detect several such stars one needs the limit of at least 24\,mag.
With a decade-long monitoring one can also detect the rotation of the lensed images. The detection of elongated images will not be possible, because this would require a nearly perfect source-lens alignment.
JWST will likely be limited by the confusion caused by stars near the Galactic center. The detection of such lensed images will provide a very clean test of general relativity, when combined with the SMBH mass measurement from orbital motions of stars, and accurate measurements of the SMBH properties, because both the source and the lens can be considered point-like. 
\end{abstract}

\keywords{
Strong gravitational lensing (1643) -- 
Supermassive black holes (1663) --
Galactic center (565) --
Optical telescopes (1174)
}

\section{Introduction}

Gravitational lensing is an important prediction of general relativity, providing both the test of the theory and a tool to detect faint but amplified sources and to measure masses also for lenses not detected by other means. The first confirmation of this effect came from the measurement of the displacement of star positions close to the Solar limb during a Solar eclipse \citep{dyson20}. Then, numerous examples of strong gravitational lensing with multiple images, arcs and rings were reported for lensed galaxies. In these cases the lenses and sources are extended and complex objects. For some applications, for example testing of general relativity, a point source lensed by a point-like lens (as already described by \citealt{chwolson24} and \citealt{einstein36}) would be more advantageous. However, until now only one star gravitationally lensed by another star has been resolved into  lensed images 
\citep{dong19}.
This is because for a stellar lens (or a less massive object) the Einstein radius is too small to be resolved with current instrumentation, a problem already noted by \citet{einstein36}. 
Lensing caused by stellar- and planetary-mass objects was detected many times with unresolved but amplified images of background stars, a phenomena called microlensing \citep{paczynski86b,paczynski96,udalski93,udalski05,wyrzykowski15,wyrzykowski20,mroz17,mroz18}.

For a galaxy lens a background star is too faint to be detected and distinguished from its host galaxy. Indeed, this was successful only for four stars (at $z=0.5$-$2$) lensed by clusters of galaxies or individual galaxies, the Refsdal supernova lensed into an Einstein cross and a delayed fifth image, predicted before its appearance \citep{kelly15,kelly16,diego16,grillo16,jauzac16}, two other supernovae \citep{goobar17,rodney21}, and a blue supergiant with a possible detection of a secondary image \citep{kelly18}.

A possibility to resolve a lensed image of an individual star is offered by the supermassive black hole (SMBH) in the center of the Milky Way, because it is massive and close enough for the Einstein ring to be resolved for some stars behind it. From the motion of stars around the SMBH, its mass was measured to be $4\times10^6\,\msun$ \citep{genzel96,genzel97,genzel00,eckart96,eckart97,eckart02,ghez98,ghez00,ghez05,ghez08}. However, even with this mass, the lensing of the orbiting stars is difficult to detect. The star S87 has the widest orbit explored so far with the semi-major axis of $2.74"$ \citep{gillessen17}. This corresponds to a maximum distance behind the SMBH of 0.05\,pc or 10\,000\,AU. At such a distance the Einstein radius is only 5\,milli-arcsec, an order of magnitude smaller than the best resolution currently achievable in the optical for stars as faint as those orbiting the SMBH. Moreover, detecting lensing signatures would require a near perfect alignment of a star with the direction towards the SMBH.

The future telescopes will reach resolution sufficient to resolve such small lensing signatures. \citet{bozza09} predicted that in 2062 the secondary lensed images of stars S6 and S27 will reach 20-22\,mag at a separation of 0.3--0.4 milli-arcsec from the shadow of the SMBH \citep[see also][]{wardle92,jaroszynski98,depaolis03,bozza04,bozza05,binnun10}. In 2047 the secondary image of star S14 will reach 23.5\,mag 0.14 milli-arcsec away. \citet{bozza12} also considered the possibility of detecting astrometric shifts of the primary images of orbiting stars due to lensing, with expected values of up to 0.3 milli-arcsec. There is also a possibility of detecting the amplification of stars behind the SMBH, but at the current sensitivity limit these events are rarer than one per century \citep{alexander99}.

The lensing signatures should however be stronger and easier to detect for background stars. The Einstein radius is $0.7$, $1.4$, and $1.6"$ for a star 1, 8, and 16\,kpc behind the SMBH, respectively. The major limitation is therefore not the resolution directly, but confusion caused by stars in the Galactic center and the number density of distant stars which could align almost perfectly with the SMBH and which could be lensed. 
\citet{chaname01} predicted microlensing events by the SMBH of several stars down to the magnitude of 21--23\,mag (see also \citealt{alexander99} for microlensing rates at lower sensitivities without separating the images)\footnote{Lensing of stars by the SMBH in the center of M31 was also considered by \citet{bozza08}.}. However, they only considered bulge stars and their adopted limits are much shallower than what will soon be available.
The lensing of stars behind the SMBH in the Galactic center has therefore not been investigated so far in the context of future telescopes. Hence, the objective of this paper is to predict whether they will have sufficient sensitivity to detect strongly lensed stars in the background of the SMBH in the Galactic center, and what limiting magnitudes are necessary for this.

We adopt the observing wavelength of $2.2\,\micron$ ($K$-band) and the Vega magnitude system, so we convert AB magnitudes by subtracting 1.85 \citep{blanton07}.


\begin{figure*}
\includegraphics[width=0.85\textwidth,clip]{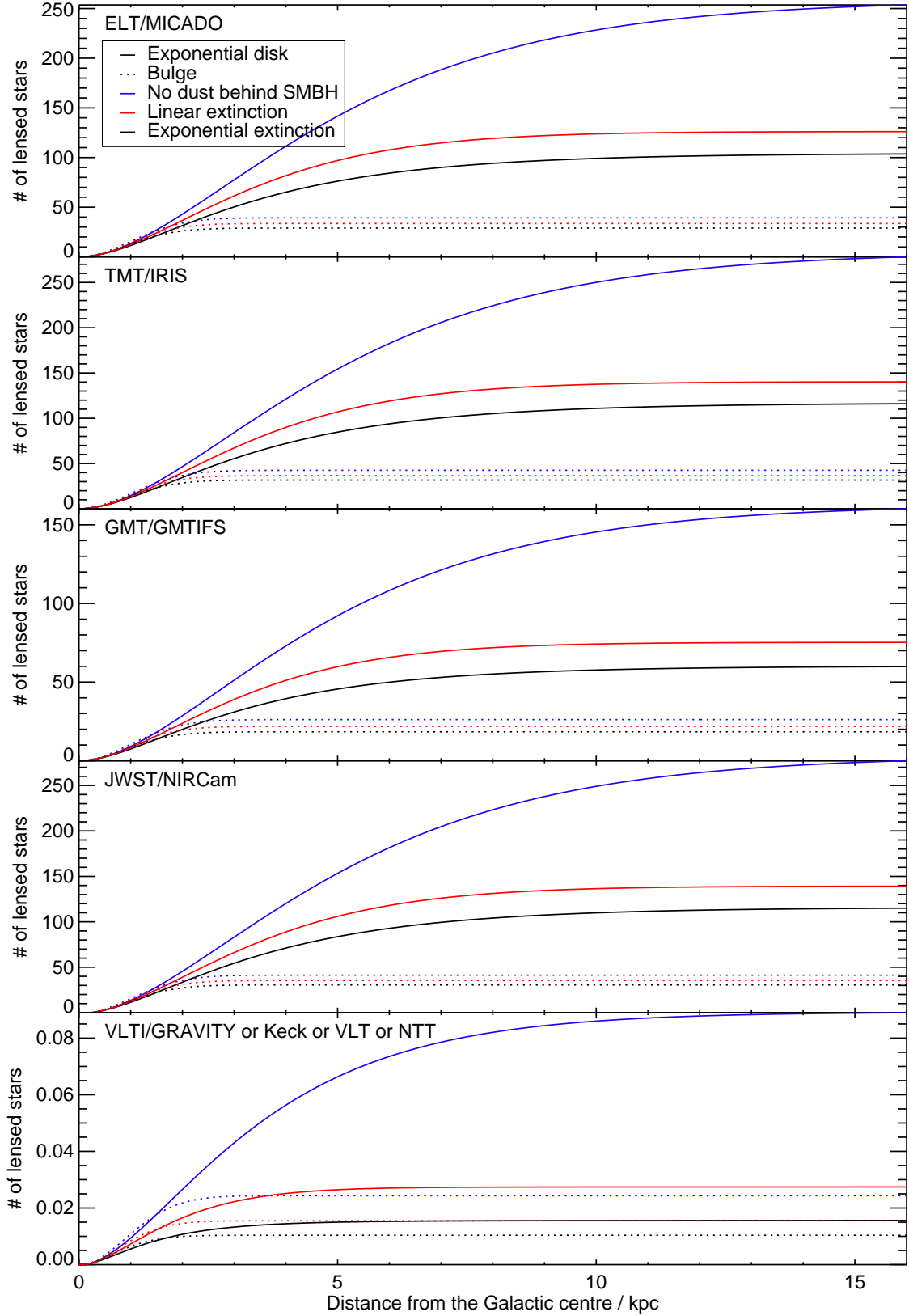} 
\caption{Cumulative number of resolved lensed images of background stars expected to be seen by high-resolution telescopes assuming $5\sigma$ limits in 5\,hr of integration (Table~\ref{tab:tel}). Solid lines correspond to an exponential disk, whereas dotted lines correspond to the bulge. Blue curves correspond to an optimistic case that the stars will be attenuated in the same way as stars orbiting the SMBH, red curves correspond to dust extinction rising linearly with distance, and  black curves correspond to the extinction model in which the dust density decreases exponentially with the distance from the Galactic center (eq.~\ref{eq:ak} in the appendix). 
}
\label{fig:n_d}
\end{figure*}

\begin{table*}
    \centering
     \caption{Assumed spatial resolutions and limiting magnitudes (Vega, $5\sigma$) in 1\,hr and 5\,hr at 2.2\,micron.}
    \begin{tabular}{cccl}
    \hline\hline
    Telescope/instrument & Resolution (milli-arcsec) & Sensitivity (mag) & References\\
    \hline
    ELT/MICADO & 11 & 27.2/28.0 & \citet{micado10,micado18}\tablenotemark{1}  \\
    TMT/IRIS & 15 & 27.3/28.2 & \citet{iris20}, \citet{iris10}\tablenotemark{2}\\
    GMT/GMTIFS & 22 & 26.2/27.1 & \citet{gmt12}, \citet{gmtifs12}\tablenotemark{3}\\
    JWST/NIRCam & 71 & 27.3/28.2 & \citet{jwst14}\tablenotemark{4}
    \citet{jwst16}\tablenotemark{5} \\
    VLTI/GRAVITY & 3 & 19.0\tablenotemark{6} & \citet{gravity17} \\ & & & \citet{bozza12}\\
    Keck or VLT or NTT  & 50 & 19.0\tablenotemark{6} & \citet{ghez08}\\
    \hline
    \end{tabular}
    \tablenotetext{1}{\url{simcado.readthedocs.io/en/latest/index.html}}
    \tablenotetext{2}{\url{www.tmt.org/etc/iris}}
    \tablenotetext{3}{\url{www.mso.anu.edu.au/gmtifs/Performance/GMTIFS-Imager-ETC.html}}
    \tablenotetext{4}{\url{jwst-docs.stsci.edu/near-infrared-camera/nircam-predicted-performance/nircam-point-spread-functions}}
    \tablenotetext{5}{\url{jwst-docs.stsci.edu/near-infrared-camera/nircam-predicted-performance/nircam-imaging-sensitivity}}
    \tablenotetext{6}{The most optimistic limit for this instrument.}
    \label{tab:tel}
\end{table*}

\begin{figure*}
\includegraphics[width=0.45\textwidth,clip]{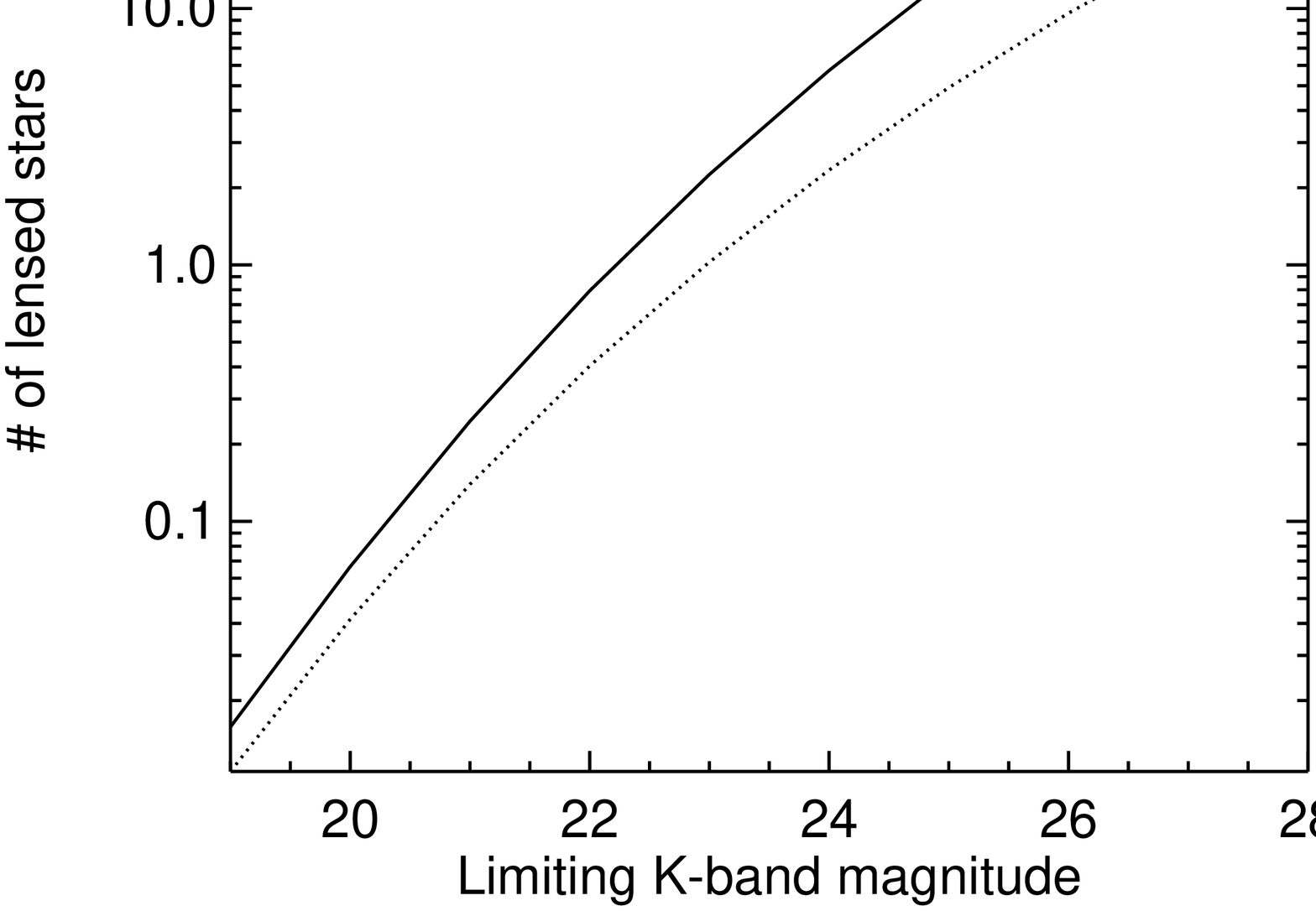} 
\caption{Number of resolved lensed images of background stars expected to be seen by high-resolution telescopes as a function of $K$-band limiting magnitude. Solid line corresponds to an exponential disk, whereas dotted line corresponds to the bulge. The extinction model in which the dust density decreases exponentially with the distance from the Galactic center was assumed (eq.~\ref{eq:ak} in the appendix). 
}
\label{fig:Nlim}
\end{figure*}


\section{Methods}
\label{sec:methods}

A star lensed by the SMBH can be identified if the secondary image on the other side of the SMBH is bright enough.
The detectability of the secondary image depends on the intrinsic (non-lensed) brightness of the star, the distance of the star behind the SMBH and the angular separation of the true image ($\beta$). Hence, we calculated the total number of detectable lensed stars by integrating along the distance from the SMBH and along the angular separation in the following steps.

\begin{enumerate}
    \item For each distance we calculated the angular size of the Einstein radius ($\er$) assuming the mass of the SMBH of $4.261\times10^6\,\msun$ and the observer-lens distance of $\dl=8.247$\,kpc \citep{gravity20}. 
    \item We considered only the distances from the SMBH for which the Einstein ring is larger by a factor of 5 than the resolution of a given telescope, so that the separation of the lensed images can be measured. This assumption has little impact on the calculations, because it excludes only very small distances and hence a very small volume. We integrated up to the lens-source distance $\dls=16$\,kpc from the SMBH, which we assume to be the radius of the Milky Way disk.
    \item For a given distance from the observer to the star $\ds=\dl+\dls$ and for a given limiting magnitude of the telescope, we calculated the absolute magnitude above which a secondary image can be detected.
    \item We corrected this absolute magnitude for dust extinction in three ways. First, for each star we applied the same correction as for stars orbiting the SMBH of $\aksmbh=2.42$\,mag \citep{fritz11}. This effectively assumes that there is no dust behind the SMBH, so is a lower limit on dust extinction. Second, we assumed that extinction is linear with distance, so for a star 8.247\,kpc behind the SMBH the extinction is twice the value measured for the SMBH surroundings. Given that most of the dust along this line-of-sight is located close to the Galactic center, this likely overestimates the extinction for large distances.
    Finally, we assumed a more realistic model with dust density decreasing exponentially with the distance from the Galactic center, and the extinction proportional to the integral of this density (eq.~\ref{eq:ak} in the appendix). \label{item:dust}
    \item For each angular separation of the true position of the star and SMBH $\beta$, we calculated the amplification factor of the secondary image $\mu_-=(u^2+2)/(2u\sqrt{u^2+4})-0.5$, where $u=\beta/\er$ \citep{schneider92}. The absolute magnitude limit from the previous point was corrected for this amplification. We considered angular separations up to $5\er$ at which the secondary image has the flux of only 0.14\% of the un-lensed star. Most of the detectable lensed stars are located within $2\er$, but such wider angles resulted in non-negligible increase for deep limits. \label{item:ampl}
    \item For each angular separation $\beta$ and for a given distance $\ds$ between the observer and the star, we calculated the volume between $\beta$ and $\beta+d\beta$ and between $\ds$ and $\ds+d\ds$ as $2\pi\ds^2\beta\, d\beta\, d\ds$. 
    \label{item:volume}
    \item In order to estimate the volume number density of stars we 
    used the star luminosity function as measured 
    within 100\,pc in the {\it Gaia} $G$ filter \citep{gaia21}. Each $G$-band absolute magnitude was converted to $K$-band using main-sequence colors \citep{pecaut12,pecaut13}\footnote{\url{www.pas.rochester.edu/$\sim$emamajek/EEM\_dwarf\_UBVIJHK\_colors\_Teff.txt}}. The result does not change significantly if we use a luminosity function measured directly in the $K$-band \citep[eq.~3 and table 2 in][]{mamon82}.   
    Then we assumed that for stars in the disk this luminosity function only applies close to the Solar neighborhood and on the other side of the SMBH with $\dls=\dl=8.247$\,kpc and scaled the number density at other distances to reflect an exponential disk, i.e.~by a factor $\exp[-(\dls-\dl)/h_{\rm disk}]$ with a scale length of $h_{\rm disk}=2.75$\,kpc \citep{zheng01}.
    We then considered the bulge applying the model of \citet{dwek95}, in which the star number density is proportional to $\exp[-0.5(\dls/\mbox{kpc})^2]$. The normalization was set so that at the Galaxy center the bulge has the number density a factor of $1.23/1.07=1.15$ higher than the disk \citep[table~2 of][]{batista11}. \label{item:model}
    \item To obtain a local number density of detectable stars at a given distance from the SMBH and at a given angular separation, we integrated this (scaled) star luminosity function above the limiting absolute magnitude derived in \ref{item:ampl}.
    \item We multiplied this number density by the volume element derived in \ref{item:volume} and obtained the number of detectable secondary images in this element.
    \item We added the contributions of each angular separation and each distance to derive the cumulative number of detectable lensed stars as a function of distance from the SMBH.
\end{enumerate}

We made these calculations for 
the Extremely Large Telescope (ELT) with the Multi-Adaptive Optics Imaging Camera for Deep Observations (MICADO),
Thirty Meter Telescope with the Infrared Imaging Spectrograph (IRIS),
Giant Magellan Telescope (GMT) with GMT Integral-Field Spectrograph (GMTIFS),
James Webb Space Telescope (JWST) with Near Infrared Camera (NIRCam),
and Very Large Telescope Interferometer (VLTI) with GRAVITY. We also made the calculations for the parameters corresponding to the best existing images taken with the Keck Telescope, Very Large Telescope (VLT), or New Technology Telescope (NTT).
The angular resolutions and $5\sigma$ limiting magnitudes in 1 and 5\,hr hours of integration used in the calculations are shown in Table~\ref{tab:tel}. For ELT, TMT, and GMT the resolution reflects the diffraction limit at $2.2\,\micron$.
For 4--8\,m-class telescopes and for VLTI/GRAVITY, instead of 1 and 5\,hr limits we used the most optimistic limit in order to show the maximum number of detectable lensed stars.

\section{Results}
\label{sec:results}

For each telescope listed in Table~\ref{tab:tel}, Fig.~\ref{fig:n_d} presents six cumulative numbers of lensed stars with detectable secondary images, corresponding to three choices of the extinction model (point \ref{item:dust} above), and two choices of the number density model, the disk and bulge (point \ref{item:model} above). We assumed $5\sigma$ limits in 5\,hr of integration for this figure, and in 1\,hr of integration for Fig.~\ref{fig:n_d_1hr} in the appendix. 
For 4--8\,m-class telescopes and for VLTI/GRAVITY we show a single panel, because we adopt the same (most optimistic) limiting magnitude.
ELT, TMT, GMT, and JWST should detect lensed secondary images of around 100 (60) stars in the disk and 30 (20) stars in the bulge, down to 28 (27)\,mag (Vega) $5\sigma$ limits requiring 5 (1)\,hr of integration. The number depends mostly on adopted extinction, because it affects mostly larger distances at which the probed volume is larger. Most of the detected stars will be located within 5\,kpc behind the SMBH, because of the decreasing number density of stars in the disk.

VLTI/GRAVITY will not be able to see any secondary images of lensed stars, as their expected number is less than 0.1. Similarly, we predict that existing 4-10\,m telescopes should not have detected any of such images, consistently with the lack of such discovery using existing images.

Fig.~\ref{fig:Nlim} shows the number of detectable secondary images of lensed stars as a function of the limiting magnitude. In order to detect several such stars one needs the limit of at least 24\,mag.

\section{Discussion}
\label{sec:discussion}

We show that with future telescopes it will be possible to look for lensed images of background stars. One would need to identify two sources exactly on the other side of the position of the SMBH, whose  spectra and/or colours are identical and whose positions and fluxes satisfy the lensing equation.

The calculations show that sensitivity is a much more important factor to resolve lensed images of background stars than resolution (though sensitivity depends on resolution indirectly due to confusion, see below). Indeed, VLTI/GRAVITY has better resolution than 20--40\,m-class telescopes, but only the latter will be able to detect secondary images of lensed stars, because they will be 8--9 magnitudes more sensitive. Similarly, JWST/NIRCam has a resolution comparable to the existing images with the Keck telescope, but we predict it to be able to detect secondary images of lensed background  stars, due to its improved sensitivity (but see the confusion limitations below). 

This is because all considered telescopes do have sufficient resolution to resolve lensed image separation, which is of the order of the Einstein radius. For a background star 1--16\,kpc behind the SMBH it is $0.7$--$1.6"$. The limitation is therefore the number of stars which are within such a small angular distance from the SMBH. In the vicinity of the SMBH the number density of stars is so high that several tens of such stars have been detected in the central arcsec; though the lensing signature for them is extremely small. For distances far enough so that the Einstein radius is larger, the number density of stars in the disk is very low. Hence, in order to detect any of them, a very deep image is required in order to probe intrinsically fainter, and hence more numerous, stars.

In these calculations we do not take into account confusion, which in principle is a limiting factor of this analysis. Most of the contribution to confusion will come from stars around the SMBH, because they are more numerous than those in the background or foreground. There are 57 stars within the central arcsec from the SMBH detected down to 19\,mag \citep[fig.~1 of][]{gillessen17}. In order to estimate the number of fainter stars in this population, first we calculated the absolute magnitudes at the distance of 8.247\,kpc (around the SMBH) for 19\,mag and the limiting magnitude listed in Table~\ref{tab:tel} and corrected them for dust extinction of $\aksmbh=2.42$\,mag, as above. Then we integrated the luminosity function,
obtaining 70 times more stars down to the apparent magnitude of 27--28\,mag than to 19\,mag, i.e.~around 4000 stars brighter than 27--28\,mag within the central arcsec. This corresponds to $0.6$ stars per ELT beam, $1.0$ stars per TMT beam, $2.2$ stars per GMT beam, and $23$ stars per JWST beam (see Table~\ref{tab:tel}). This means that due to their unprecedented resolution, ELT, TMT, and GMT will only start to be confused at the listed limiting magnitudes towards the Galactic center. JWST will be heavily confused towards the Galactic center, so the possibility to detect lensed stars in the background will rely on the luck of the existence of a bright star in the background and accurate subtraction of stars around the SMBH. This may also turn out to be possible if the initial mass function in the Galactic center is much more top-heavy, with much fewer faint stars than we calculated above. Otherwise, only the limit of 20\,mag will result in one source near the Galactic center per JWST beam. However, at this shallower limit only up to 0.6 lensed background star is expected.

The detection of elongated arcs of lensed background stars will not be possible with any telescope. Taking an optimistic case of a giant with a radius ten times larger than that of the Sun only 0.5\,kpc behind the SMBH results in an angular size of 10 micro-arcsec. Hence, in order to detect the elongation of the image with the best future resolution, the star would need to be stretched by a factor of at least 1000. This requires an extremely accurate alignment of the star with the line towards the SMBH. The tangential magnification for a case of a point lens can be expressed as $\mu_t = x^2 / (x^2 - 1)$, where $x=\theta/\er$ and $\theta$ is the angular separation of the image and the lens \citep{schneider92}. Hence, $\mu_t=1000$ corresponds to $\theta=1.001\er$, so the image forms almost at the Einstein radius. Using the lens equation for a point lens $\beta=\theta-\er^2/\theta$, one obtains the true source separation from the lens of only $\beta=0.002\er$. Considering only the number of stars in such a narrow cone, one gets their total number a factor of $1000^2$ lower than for stars within $2\er$, calculated above and hence there will be no stars that close to the SMBH. The increased lensing amplification of the primary image will not help here, because at such small separations from the lens it is almost equal to the amplification of the secondary image, already considered in these calculations.

In order to look for or to confirm a pair of objects as lensed images, one can try to detect their rotation around the SMBH due to relative motion of the source and lens. Proper motion in units of milli-arcsec per year can be calculated from the linear velocity $v$ in {\kms} and distance $D$ in kpc as $v/(4.74D)$. Due to the orbital motion of the Sun around the Galactic center, the SMBH moves with respect to the Sun with a linear velocity of $200\,\kms$, whereas a background stars moves with a linear velocity of $400\,\kms$, assuming a flat rotation curve. Therefore for a star $\dls=1$, $3$, $10$, $16$\,kpc behind the SMBH, its relative proper motion with the SMBH is $\mu=400/(4.74\ds) - 200/(4.74\times\dl)=4$, $2.4$, $0.5$, and $1.6$ milli-arcsec per year, respectively. The Einstein radius crossing time (Einstein timescale) is $t_E\equiv\er/\mu=170$, $440$, $3100$, and $1000$ years. Hence, especially for lensed stars not too far behind the SMBH, the comparison of images taken around a decade apart should reveal the rotation of the lensed images.

\section{Conclusions}
\label{sec:conclusion}

We show that  ELT, TMT, and GMT will resolve lensed images of around 100 (60) stars in the disk and 30 (20) stars in the bulge in the background of the SMBH down to 28 (27)\,mag (Vega) limits at $K$-band, requiring 5 (1)\,hr of integration. 
In order to detect several such stars one needs the limit of at least 24\,mag.
With a decade-long monitoring one can also detect the rotation of the lensed images. JWST will likely be limited by the confusion caused by stars near the Galactic center. If these observations are successful, then together with the SMBH mass measurement from orbital motions of stars, this will provide a very clean test of general relativity, because both the source and the lens can be considered point-like. For such test the contribution of stars to the lensing signal will need to be taken into account. Within the central arcsec (0.2\,pc) the mass contribution of the SMBH is around 35 times higher than that of stars \citep{genzel97}, so this is not problematic, but may hide subtle deviations from general relativity. Moreover, the exact position and mass of the SMBH will be measured in an independent way. This will provide a cross-validation of measurements using orbital motions of stars. The effect of the SMBH spin will be of the order of 4 micro-arcsec for the first order \citep{sereno06}, so this will not be constrained by telescopes considered here.
The detection of stretched images of the lensed stars will not be possible, because given extremely small angular sizes of stars, the alignment with the SMBH would need to be better than $0.002\er$. The probability that a star happens to be in such a small volume is almost zero.

\acknowledgments 

We wish to thank the referee for useful suggestions, and Joanna Baradziej, Mattia Negrello, Jean Surdej, and 
\L ukasz Wyrzykowski for discussions and comments.
MJM acknowledges the discussion with Hong Du on gravitational lensing, which was a prompt to make calculations presented in the Letter.
MJM acknowledges the support of 
the National Science Centre, Poland through the SONATA BIS grant 2018/30/E/ST9/00208.
This research has made use of 
the NASA/IPAC Extragalactic Database (NED) which is operated by the Jet Propulsion Laboratory, California Institute of Technology, under contract with the National Aeronautics and Space Administration;
the WebPlotDigitizer of Ankit Rohatgi ({\tt arohatgi.info/WebPlotDigitizer})
and NASA's Astrophysics Data System Bibliographic Services.



\appendix

\section{Derivation of the extinction profile from exponential distribution of dust}

\begin{figure}
\includegraphics[angle=0,width=0.45\textwidth,clip]{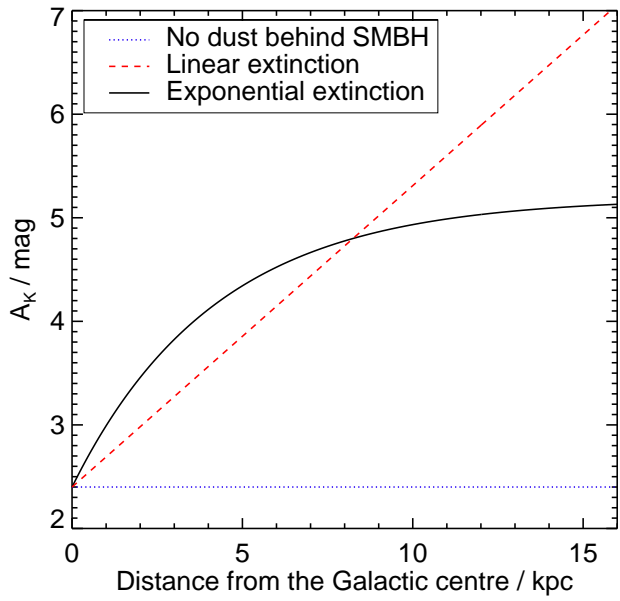} 
\caption{Extinction as a function of distance from the SMBH using the exponential model (black solid line; eq.~\ref{eq:ak}), the linear model (red dashed line) and the constant model in which extinction for all stars behind the SMBH is assumed to be the same as that measured for stars orbiting the SMBH (blue dotted line). By the symmetry requirement, 8.247\,kpc behind the SMBH the exponential and linear models result in the extinction twice as that around the SMBH.
}
\label{fig:ak_d}
\end{figure}

We assumed the exponential Galactic dust density ($\rho$) model from \citet[][eq.~16]{sharma11}, which in the plane of the disk can be expressed as $\rho=C\exp(-\dls/h)$, where $C$ is a normalization constant, $\dls$ is the distance from the Galactic center (consistently with the notation above) and $h=4.2$\,kpc is the dust distribution scale length. The extinction for a star at a distance {\dls} behind the SMBH is therefore the extinction measured for stars close to the Galactic center plus the integral of this dust density
\begin{equation}
A_K=\aksmbh + \int_0^{\dls} \rho\, d\dls =
\aksmbh + C\left[-h\exp\left(-\frac{\dls}{h}\right)\right]_0^{\dls}=
\aksmbh +Ch\left[1-\exp\left(-\frac{\dls}{h}\right)\right]
\label{eq:akwithc}
\end{equation}
The value of the constant $C$ can be obtained from the requirement of symmetry that $8.247$\,kpc behind the SMBH the extinction should be twice as that measured for the Galactic center, i.e.~setting $\dls=\dl$ and $A_K=2\aksmbh$ one obtains
\begin{equation}
C=\frac{\aksmbh}{h\left[1-\exp\left(-\frac{\dl}{h}\right)\right]}    
\end{equation}
and substituting this in eq.~\ref{eq:akwithc}, one obtains
\begin{equation}
A_K=\aksmbh\left[1+\frac{1-\exp\left(-\frac{\dls}{h}\right)}{1-\exp\left(-\frac{\dl}{h}\right)}\right] 
\label{eq:ak}
\end{equation}
This is used for the extinction model in section \ref{sec:methods}, point \ref{item:dust} and is compared with the linear and constant models on Fig.~\ref{fig:ak_d}.

\section{Number of lensed stars with 1\,hr limiting magnitudes}

\begin{figure*}
\includegraphics[width=0.85\textwidth,clip]{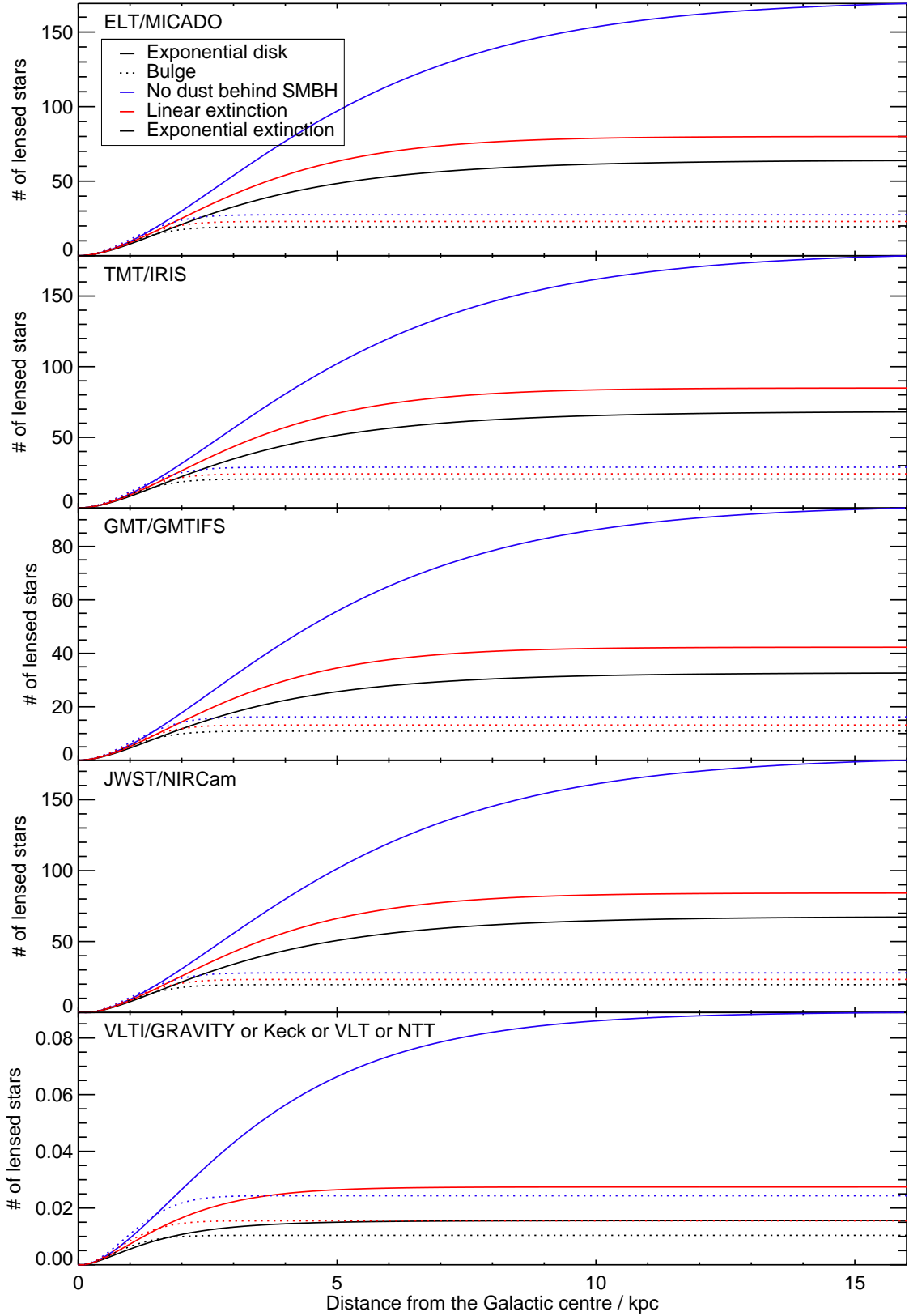} 
\caption{The same as Fig.~\ref{fig:n_d}, but for assuming $5\sigma$ limits in 1\,hr of integration (Table~\ref{tab:tel}).
}
\label{fig:n_d_1hr}
\end{figure*}

\end{document}